\documentclass{aa}  
\usepackage{graphicx}
\usepackage{natbib}
\begin{document}

\title{On the reliability of the fractal dimension measure of solar magnetic features and on its variation with 
solar activity}


\author{S. Criscuoli \inst{1,2},  M.P. Rast \inst{3},
                      I. Ermolli \inst{4} \and M. Centrone  \inst{4}}
   \offprints{S.Criscuoli}
   \institute{Dipartimento di Fisica, Universit\`a di Roma ``Tor Vergata'',
   Via della Ricerca Scientifica 1, I-00133 Roma, Italia
   \and
   High Altitude Observatory, NCAR, Boulder, CO 80307-3000, USA
 \email{serenac@hao.ucar.edu}
\and 
Laboratory for Atmospheric and Space Physics, Department of Astrophysical and Planetary Sciences, University of Colorado, Boulder CO 80309, USA
\and   
INAF - Osservatorio astronomico di Roma , Via Frascati 33, I-00040, Monte Porzio Catone, Italia
}
\authorrunning{S. Criscuoli et al.}
\titlerunning{On the reliability of the fractal dimension measure of solar magnetic features}

\date{Received ; accepted }
 
\abstract
{Several studies have investigated the fractal and multifractal nature of magnetic features
in the solar photosphere and its
variation with the solar magnetic activity cycle.}
{Here we extend those studies by examining the fractal geometry of bright 
magnetic features at higher atmospheric levels, specifically in the solar chromosphere.  
We analyze structures identified in CaIIK images obtained with the Precision Solar Photometric 
Telescopes (PSPTs) at Osservatorio Astronomico di Roma (OAR) and Mauna Loa Solar Observatory (MLSO).}  
{Fractal dimension estimates depend on the estimator employed,
the quality of the images, and the structure identification techniques used. 
We examine both real and simulated data and employ two 
different perimeter-area estimators in order to understand the sensitivity of the deduced fractal 
properties to pixelization and image quality.}
{The fractal dimension of bright 'magnetic' features in CaIIK images
ranges between values of 1.2 and 1.7 for small and large structures respectively.
This size dependency largely reflects the 
importance of image pixelization in the measurement of small objects.
The fractal dimension of chromospheric features 
does not show any clear systematic variation with time 
over the period examined,
the descending phase 
of solar cycle 23.}
{These conclusions, and the analysis of both real and synthetic images on which they are based, 
are important in the interpretation of previously reported results.}
\maketitle

\section{Introduction}
The spatial distribution of the solar magnetic fields is very complex, depending on both
the varying level of magnetic activity over
the solar cycle and the height of observation in the solar atmosphere.
This complexity likely results from both
the dynamo process itself, which may occur on many different 
spatial scales, and the interaction of the field with 
convective motions as it emerges through the Sun's outer layers.
The signatures of these processes have been investigated in previous works 
by fractal analyses 
of solar active regions, but the 
quantitative results obtained 
differ widely depending on the type of data and analysis 
techniques employed 
\citep[e.g.][]{Jansen2003,McAteer2005}.  
Moreover, the lack of a unique 
definition of the fractal dimension itself often 
makes comparison of results difficult.

Among recent studies, one focused on the
possible relationship between magnetic feature complexity and solar cycle phase 
\citep{Meunier2004}. This study looked at a large sample of
active region magnetograms 
acquired with the Michelson Doppler Interferometer (MDI) aboard
the Solar and Heliospheric Observatory (SOHO)
between April 1996 and June 2002. 
The measured fractal dimension increased with structure 
size (in agreement with \citet{Meunier1999} and 
\citet{nesme96}), showing a 
peculiar change in behavior near structures of area
550-800 ${\rm Mm}^{2}$. A similar dependence on structure size was also found by \citet{Jansen2003}
in both high resolution photospheric magnetograms and magnetohydrodynamic (MHD)
simulations. Additionally, \citet{Meunier2004}
investigated the relationship between the geometry of facular structures of
different spatial scales and magnetic field intensity,
flare activity, and solar cycle phase. 
She found that, while a region complexity generally increases with 
magnetic field intensity, there is no clear
correlation with flare activity. Variations of fractal dimension with solar cycle were also reported,
but their amplitudes and sign largely depended on object size and the associated magnetic field.

In order to investigate the complexity of magnetic features 
using observations representative of 
chromospheric heights, 
we have analyzed the fractal dimension of bright features 
identified in full-disk 
CaIIK images acquired by 
the Precision Solar Photometric Telescopes (PSPTs) at Osservatorio Astronomico di Roma (OAR) and Mauna Loa Solar Observatory (MLSO).
The data analyzed span the past 6 years and thus allow investigations of 
variation with the solar cycle. 

Several factors can influence fractal dimension estimation. 
Both image resolution and projection effect correlation, mass 
function, and perimeter-area estimators in studies of 
interstellar molecular clouds \citep{Sanchez2005,Vogelaar1994},
and resolution and thresholding effects
are important in fractal dimension estimation of flow 
patterns in field soil by box-counting methods
\citep{Baveye1998}.
\citet{Lawrence1996} studied similar effects in multifractal and fractal measures
(box-counting, cluster dimension, threshold set) of
solar magnetic active regions. The effect of structure selection technique
was also investigated by \citet{Meunier1999,Meunier2004}.

To link our findings directly to
those of several recent solar studies \citep[for example][]{Meunier2004,Jansen2003} 
we have employed the  perimeter-area relationship
to define the fractal dimension of the identified features.
To interpret our results, we have investigate the sensitivity of the 
deduced fractal dimension
to the pixelization and resolution of the image 
and to the perimeter measure algorithm employed.
In particular, we have determined 
how these factors influence the geometric properties deduced for
objects of different sizes, and more generally have addressed
the question of whether the perimeter-area relation is suitable 
to the study of the fractal and multifractal nature of 
solar magnetic features as a function of the solar cycle. 

This paper is organized as follows.
In the next section we 
describe the observations, data processing techniques, and geometrical measures
employed.
In section \S3 we present the results obtained 
and in \S4  compare them
to those of previous efforts.
In \S5 we investigate the sensitivity of the deduced fractal dimension 
to the image resolution and the measurement techniques employed,
by examining synthetic structures whose 
fractal properties are theoretically known: non-fractal objects,
von Koch snowflakes, and those produced by fractional Brownian motion (fBm).
The effect of atmospheric seeing is also investigated using PSPT images taken under variable seeing
conditions. 
In section \S6 we discuss our and previous results,
in light of the conclusions drawn in \S5. 
Finally, in \S7 we summarize our work and conclude 
with a more general discussion on the validity of the methods.
\section{Observations, processing and definitions}
\subsection{PSPT data} 
The bulk of the data we analyzed is from the
archive of daily full-disk observations 
carried out with
the PSPT at OAR. This was supplemented with data from the PSPT at 
MLSO for consistency and resolution tests (\S3.1 and \S5.2).
Details about the  data and the image pre-processing 
can be found in \citet{ermolli06}. In brief, 
the images were taken with ''twin'' telescopes at the two sites, 
through interference filters centered at three wavelength bands 
(CaIIK line center 393.4nm, fwhm 0.27nm, blue continuum 409.4nm, fwhm 0.27nm, 
and red continuum 607.1nm, fwhm 0.46nm), 
with a 2048 $\times$ 2048 16 bit/pixel CCD camera, yielding
a spatial scale of $\sim 1\arcsec$ per pixel. 
Images from OAR were 
binned to half resolution, yielding
a final spatial scale of $\sim 2\arcsec$ per pixel. 
Images from the two telescopes were independently 
dark and flat-field corrected and had the
mean center-to-limb variation removed. The images of any one wavelength triplet were resized and aligned to
allow pixel by pixel comparison between filters.

For this study we selected OAR daily image triplets, obtained on 238 different 
observing days during the summers (July to September) of
2000 through 2005. We chose images acquired during the summer months because
these are generally of higher quality.
In order to compare results obtained with the two instruments, we also selected 44 triplets 
(the best in the CaIIK band of the day, according to quality criteria described in
\S2.2) from the MLSO and OAR archive taken 
during the summer of 2005. For that comparison, MLSO images were rescaled to match OAR spatial 
scale images ($\sim 2\arcsec$ per pixel).

Finally we were able to quantify the effects of atmospheric seeing by using MLSO images acquired at 
10 minutes intervals throughout the 
day, weather permitting. According to the quality criteria explained in next paragraph,
we selected 27 pairs of  \emph{high} and \emph{low} quality triplets (one pair per day) from the period February 
to October 2005. For this analyses, the full resolution ($\sim 1\arcsec$ per pixel)
MLSO data were employed.

\subsection{Data quality} 
The geometric properties of solar features extracted from the images are likely 
sensitive to the spatial resolution of the image being analyzed.  This in turn depends
on  atmospheric and instrumental operation conditions  during the observation.  To estimate the inherent 
quality of any given data image, we measured (in pixel units) the width of a Gaussian fit to the 
limb profile observed in CaIIK images.   
Small values of the solar limb width indicate lower instrumental or atmospheric smearing
and thus better quality images.
 The limb width distributions of our datasets
are asymmetrically shaped with a long tail toward higher values, so that the mean is not the most probable value.
The mean limb profile width of the OAR CaIIK (binned to half resolution) images analyzed is $2.5\pm4.0$ pixels with a 
median value of $1.4$ pixels. That of the MLSO 
CaIIK images from the summer 2005 is $4.1 \pm 0.8$ pixels for the mean and $3.9$ pixels for the median (measured on full 
resolution images), while that 
for the OAR images acquired in the same period are $4.0\pm6.5$ pixels and $2.2$ pixels respectively (measured on binned half  
resolution images).
Considering the different pixel scale of MLSO and OAR images, the two 2005 datasets have similar median quality
(about 4 arcsec), but the OAR dataset contains a higher number of low quality images, skewing the mean to a much 
higher value.

\begin{figure}
\centering
\begin{tabular}{c}
\includegraphics[width=6.5cm]{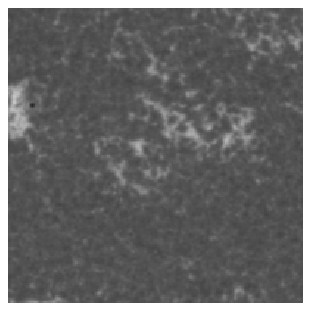} \\   
\includegraphics[width=6.5cm]{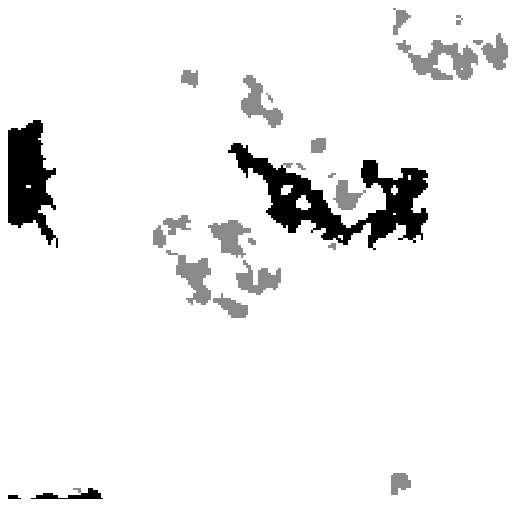}\\ 
\end{tabular}
\caption{ Central disk detail of a CaIIK image from OAR PSPT archive and corresponding mask obtained with 
the second identification 
method described in the text. Features with areas lower and larger than 2000Mm$^{2}$ are shown in gray and black 
colors respectively.}
\label{figpsptimm}
\end{figure}

The MLSO limb width distribution for year 2005 has its maximum
at a value of about $3.7$ pixels (measure on full resolution images). 
To study fractal dimension dependence on seeing condition (\S5.2),
images from each day were grouped in two sets:
those 
whose limb widths lie below $0.1\sigma$ from the peak,
and those whose limb widths lie between 
$0.8\sigma$ and $3.0\sigma$ above the peak.
From each of these groups, the best and the worst images were selected, so that 
for each observing day, two different quality images were retained.
This restricted our analysis of seeing effects 
to 27 observing days.
The mean limb width for the two groups were $3.5\pm0.2$ and $4.9\pm0.3$ for \emph{high} and \emph{low} 
quality images respectively.

\subsection{Feature identification}
Bright features were identified in the CaIIK images using two methods
based on a combination of pixel intensity and connectivity. Because of the link 
between CaIIK brightness
and magnetic flux intensity \citep[e.g.][]{sku75,harvey1999,rast2003,ortiz_rast2005} these 
features are assumed to represent small magnetic structures, but this assumption 
plays no role in their identification.

The first identification method is analogous to that used by 
\citet{Meunier2004}, taking into account intensity alone and based on fixed 
thresholding values. While \citet{Meunier2004} employed thresholding on magnetograms,
we select pixels whose intensity contrast
\footnote{$I_c=\frac{{I-I_0}}{I_0}$, where $I$ is the intensity measured at each pixel and
$I_0$ is that representative of the quiet sun and obtained by a fit to its
center to limb intensity variation.} 
in the CaIIK images exceeds a given value. 
Umbral, penumbral and pore pixels, were 
included selecting those pixels whose intensity
in red continuum images 
was below a given threshold.

The second identification method applied takes into account both pixel intensity and connectivity. 
As described in detail by \citet{ermolli06}, we first identified regions of the solar disk 
which include 
active regions and their remnants. These pre-selected regions are made up of
those pixels which are brighter than 
a fixed 
contrast value on a spatially filtered image (box average 
$side=\frac{R_{sun}}{20}$). A second contrast threshold value, determined by the procedure explained 
in \citet{nesme96},
is then used to single out from the pre-selected regions all those features identified for our study. 
Pixels representative of sunspots and pores, previously identified from red continuum
images, were excluded from the 
identified features.  

For the subsequent fractal analysis, the two identification methods were employed to produced two independent
binary masks from each image triplet processed. 
In  each of these pixels satisfying one of the 
two identification criteria above were assigned a value of one,
with all other pixels set to zero. 
An example of a CaIIK PSPT image (showing only a central disk region) and the corresponding mask, obtained with the second identification method described,
is given in fig.\ref{figpsptimm}. 
To reduce distortion due to projection effects, the analysis was restricted to structures near disk center, $\mu > 0.8$, 
where $\mu$ is the cosine of the heliocentric position angle. Additionally, isolated 
bright points were removed from consideration by discarding all structures of area less than 
10 pixel$^{2}$. 

\subsection{Perimeter and area evaluation} 
There are several ways to define and evaluate the perimeters and areas of 
features in a binary image \citep{gonzalez2002}, and thus 
characterize the independent structures.
The goal is
to define, detect, and count the pixels which 
constitute the feature edges. For our study we considered three methods, and
evaluated the errors associated with them.

In the first method,
we defined border pixels by row and column, identifying for each the pixel 
for which the binary value changes.
The perimeter was then evaluated by summing the external sides of the border pixels, 
so that for example an object 
made up of 1 pixel has an area of 1 and a perimeter of 4, while 
one made up of two pixels has
an area of 2 and a perimeter of 6 or 8 depending on the pixels' 
relative positions. 

In the second method, we applied the Roberts operator 
\citep{fractgeomdigim} to the image 
in order to identify border pixels and defined the perimeter as the sum 
of the all pixels whose value is not zero. Using
this method, an object of 1 pixel has a perimeter of 4
and an object of two pixels always has a perimeter 6, 
independent of the relative positions. 

In the third method,
pixels are identified as border pixels if they are connected from 
between 1 and 7 of the neighboring 8 contiguous pixels. 
The perimeter is the sum of the selected pixels, so that 
an object 1 pixel in area has 
a perimeter of 1 and an object of area 2 
has a perimeter of 2, independent 
of the pixels' relative positions. 

Only the results obtained using
the first method are included in the body of this paper.  The reasons for
this are discussed in \S5.1.

\subsection{Fractal dimension definition}
Several definitions of the fractal dimension of two-dimensional structures 
and corresponding techniques for its
estimation exist \citep{fractgeomdigim}.

If a structure is self-similar,
its perimeter $L$ and area $A$ display a power-law relation:
\begin{equation}
\label{fracdimdef}
L \propto A^{d/2} \ ,
\end{equation}
where $d$ is the fractal dimension.
With this definition, $1 \leq d \leq 2$ and  
\emph{d=1} for non-fractal structures.

We estimated \emph{d} using two methods. 
In the first, we performed a simple linear fit to the logarithm of the 
perimeters and areas measured for structures of different sizes;
we indicate the fractal 
dimension so obtained as \begin{em}D\end{em}. In order to investigate the size dependence of the fractal dimension
estimated in this way, the fit is performed over the entire data set or 
over objects in a specified size range.
In the second, we adopted a method first proposed by 
\citet{nesme96} and later employed by
\citet{Meunier1999,Meunier2004}, in which 
perimeter and area values are averaged over bins in area, each of 
width $\Delta \log A=0.05$, and the fitting is done on these averages
for a series of overlapping windows of constant width $\Delta \log A=1.5$, producing a
measure of \emph{d} which is a function of \emph{A}. We indicate the
fractal dimension estimated in this way as \begin{em}d1\end{em}.
For both methods linear fits were performed by a chi square minimization and the associated error is taken to be the variance in 
the estimate of the slope.  

More details about the significance of the two estimators and how they relate to each other are given in the Appendix.  
\section{Results}
\subsection{Fractal dimension and feature size}
\begin{figure}[]
\includegraphics[width=8.0cm]{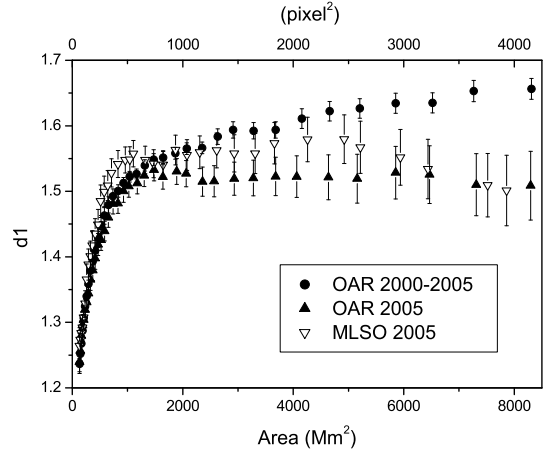}
\caption{Fractal dimension $d1$ versus area of bright features identified on calcium images ($\sim 2\arcsec/pixel$). 
Full circles: summers 2000-2005 OAR-PSPT. Open triangles: summer 2005 MLSO-PSPT. Full triangles:
summer 2005 OAR-PSPT. $d1$ increases fast with object size at area smaller than 2000 Mm$^{2}$.
For larger areas, a plateau is observed for summer 2005 OAR and MLSO data, and a slow rise on the 2000-2005 OAR
dataset. 
		 \label{figpspt}}
    \end{figure}

\begin{table} %
\centering 
\begin{tabular}{ccc}                                        
\hline
$OAR 2000-2005$  & \multicolumn{1}{c}{$OAR 2005$}    & \multicolumn{1}{c}{$MLSO 2005$}  \\
\hline
 $1.337\pm0.002$  & $1.307\pm0.003$  & $1.307\pm0.004$ \\
 $1.64 \pm 0.02$  &  $1.53 \pm 0.07$  &  $1.54 \pm 0.09$ \\
                       
\hline
\end{tabular}
\caption[]{Fractal dimension $D$ estimated for features selected on OAR-PSPT and MLSO-PSPT CAIIK images. 
First row: fractal 
dimension $D$ considering the entire range of structure areas. 
Second row: fractal dimension $D$ considering only structures 
larger than 2000 Mm$^2$. 
  }\label{psptmlsoDres}
\end{table} 

Figure \ref{figpspt} shows the variation in fractal dimension \begin{em}d1\end{em} of the identified 
chromospheric features as a function of 
their size, as derived from the OAR and MLSO PSPT CaIIK rebinned data using the second identification method 
described in \S2.3.
The results obtained from three data sets are shown: the full 2000-2005 OAR summer period, the single 2005 
summer OAR data, and the single summer 2005 MLSO data.
For all data sets, $d1$ increases with object size, increasing fastest
for structures of smallest areas and becoming almost constant at the largest
scales. The three curves overlap for structures of area less than about 1000 ${\rm Mm}^{2}$, 
corresponding to about 500 ${\rm pixel}^{2}$. For objects of size greater than about 
1000-1500 ${\rm Mm}^{2}$ the 2005 OAR and MLSO data both show a 
plateau in the measured fractal dimension. Somewhat surprisingly, this plateau is less evident
when analyzing structures from the full 2000-2005 OAR data set. The fractal dimension deduced over 
this longer period continues to slowly increase even at the largest scales.

Table \ref{psptmlsoDres} shows the value of \begin{em}D\end{em} obtained from the different data sets
when a single fit to the perimeter area relation is made over structures of all sizes (top row) or 
only those of area larger than 2000 ${\rm Mm}^{2}$ (bottom row), the threshold value suggested by the trends observed
in fig.\ref{figpspt}.
In agreement with \begin{em}d1\end{em} estimates,
the fractal dimension \begin{em}D\end{em} is reduced by the inclusion of the 
small and apparently less complex regions.
In this case, because of the large number of objects with small areas, $D$ is biased toward a low value and
the formal error quoted 
is quite small (as 
is a $\chi^2$
measure of the fit) in spite of the fact that a single linear fit does not reflect the perimeter area relation at 
all scales (see also Appendix).
 The large  
objects, who's perimeter area relationship is poorly fit by the  
single slope estimator, are insufficient in number to significantly  
alter the fit value or influence the error measure. 
Figure \ref{figpsptimm} 
displays typically structures of areas larger and smaller than 2000Mm
$^2$. The smallest objects appear somewhat rounder and more regular  
than the largest ones, but no overwhelming difference between the two  
groups can be inferred by visual inspection alone.

\subsection{Temporal variation}
\begin{figure}[]
\includegraphics[width=8.0cm]{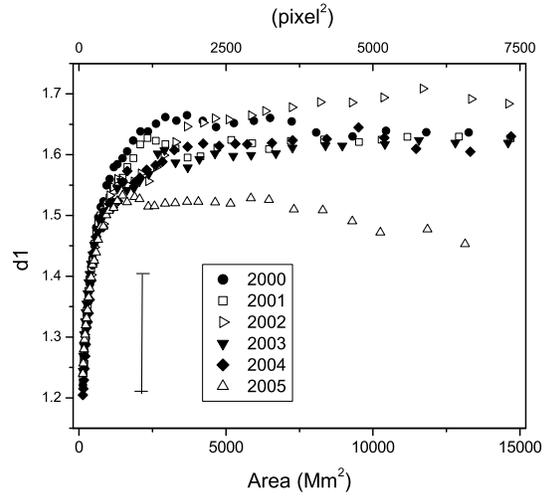}
\caption{Temporal variation of the fractal dimension $d1$ versus area for features identified on OAR-PSPT calcium images.
The bar on the left represents the largest error bar, obtained for the largest areas for year 2004. At area smaller than about
1000 Mm$^{2}$ all the curves overlap, while differences (not clearly correlated with solar cycle) are observed at the largest areas. 
		 \label{psptanni}}
    \end{figure}
\begin{figure}
\includegraphics[width=8.0cm]{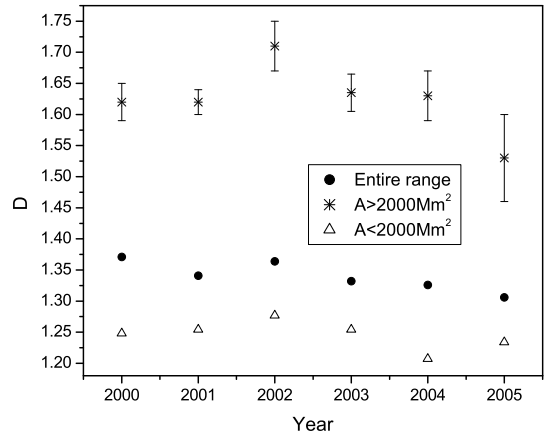}
\caption{Temporal variation of the fractal dimension $D$ versus area for selected OAR-PSPT calcium images 
and for different area range. Error bars in the case of fits performed on the whole dataset (circles) or at
smallest objects (triangles) are smaller then the symbol size. Results obtained for the largest area are in 
good agreement with results obtained
by $d1$ estimator (fig.\ref{psptanni}). 
		 \label{D00_05anni}}
    \end{figure}

Figure \ref{psptanni} shows the variation of fractal dimension \begin{em}d1\end{em} with both feature size and time 
for the six year OAR period analyzed. For sake of clarity,
only the largest error bar 
(belonging to the largest area objects of year 2004) is 
indicated on the plot. 
The fluctuations in the other values shown are generally 
smaller than the largest differences observed among 
the years.
We find that the variation in fractal dimension does not show a clear correlation with
solar cycle over the period analyzed (the descending phase of 
Solar Cycle 23). The values for large structures show 
significant year to year variation, with the maximum
and the minimum dimensions 
measured for years 2002 and 2005 respectively. 
The reliability of the 2005 values are supported by the nearly identical results obtained from the
independent OAR and MLSO measurements (Table~1, fig.\ref{figpspt}). The other years show a plateau value 
at about 1.6, although in the area range 2500-7000 ${\rm Mm}^{2}$ year 2000 has a mean value of about 1.65
and a slight increase with object size up to a maximum value of 1.7 is measured for year 2002.
Note that 
if we restrict the analyses to the area range 2500-7000 ${\rm Mm}^{2}$ then a weak trend with the last cycle, showing 
a double activity peak in 2000
and 2002, is observed. Figure \ref{D00_05anni} shows the temporal variation in
\begin{em}D\end{em} for three different area ranges. For the reasons explained in \S3.1,
error is omitted when fit is performed on the entire area range and at the smallest areas.
 In this measure a small trend with solar cycle is
observed when the entire structure size range is included in the fit. 
When fit is performed on the largest objects, the highest fractal dimension
is measured for year 2002, and the lowest for year 2005, and similar 
values are measured for the other years, in agreement with results obtained for $d1$. 
For smaller magnetic regions,
again the maximum is observed in 2002, but a minimum is found for year 2004 data. 
We notice that, while suggestive, the trends at larger objects are small and the fractal dimension is constant within the measurement
uncertainties. Moreover variations of coefficients evaluated over the entire area range and at smaller areas probably reflect
 variations in the size distribution of the magnetic regions \citep{Meunier2003, ermolli06} rather than a real 
 temporal variation of the fractal dimension. 
 
\section{Comparison to previous results}
\subsection{Fractal dimension and structure size}
Since the work of \citet{roudier87} the fractal geometry of structures found in images of 
the outer layers of the solar atmosphere has been
investigated by a number of authors.
Both magnetic features, at moderate to high
spatial resolution, and non-magnetic features 
associated with plasma motions have been studied \citep{roudier87,lawrence93,balke93,nesme96,berrilli98,
Meunier1999,stenflo03,Jansen2003,Meunier2004,McAteer2005}.
In order to ensure a meaningful comparison, we will compare the results we have found here 
only to those previously published results for active regions 
investigated using the perimeter-area estimator.

In agreement with previous results, we find a fractal dimension that increases with feature size, 
from a minimum value of about $1.2$,
to an approximately constant value of $1.5-1.7$ for structure areas larger than 
$\sim 1000 - 2000$ Mm$^{2}$. 
This range in $d1$
agrees well with measurements by 
\citet{nesme96}, but not with those reported by
\citet{Meunier1999, Meunier2004}, who found
generally a higher minimum value (around 1.4). 
From the analysis of both real and simulated data (\S 5)
we know that the 
minimum value measured 
is somewhat dependent on both the identification method employed 
and the image resolution. We suggest that the higher minimum value reported by \citet{Meunier1999, Meunier2004} 
may be a consequence of image resolution (see
\S5.3), as the full-disk MDI data she analyzed are 
unaffected by atmospheric degradation.  
The plateau in $d1$ beyond object sizes of
2000 ${\rm Mm}^{2}$ also agrees with previous results, but this time more so with those of
\citet{Meunier2004} and less so with \citet{nesme96}, who found the plateau to occur already for 
structures of size $>300$
square-pixels (corresponding to about 500 ${\rm Mm}^{2}$).
Tests with both real and simulated data suggest that this difference 
may lie in the area range over which the fit for 
$d1$ at each point was made, 
a value not quoted by \citet{nesme96}, but taken in our study to be $\Delta \log A=1.5$ in agreement
with that used by \citet{Meunier2004}.

An increase of fractal dimension with magnetic feature size was also observed by \citet{Jansen2003}. 
They studied fractal dimension of magnetic features analyzing high resolution (approximately 0.4\arcsec) magnetograms
acquired with the Vacuum Tower Telescope and synthetic 
images obtained through MHD simulations. 
They found $D=1.38\pm0.07$ on synthetic data, and $D=1.21\pm0.05$
on real data, corrected to $D=1.41\pm0.05$ when taking in to account resolution effects. Note that our estimate, 
$D\cong1.3$ (fit on entire area range), lies in between these last two values.
The
plots of \citet{Jansen2003} show a deviation from these fits for $\log$(A/pixel$^{2}$)$>$2.5 (corresponding to 315 pixel$^{2}$) 
for both real and simulated data, despite the differing pixel scale of the two data sets ($\sim$ 72 km 
for real data
and $\sim$ 21 km for simulated ones). Fits to objects whose
areas  were larger then this threshold gave $D=1.47$ 
and $D=1.9$ (values not corrected for resolution effects) for real and simulated objects respectively. 
\\

\subsection{Temporal variation}

\citet{Meunier2004} performed a time-dependent
analysis, evaluating the variation in the
fractal dimension \begin{em}d1\end{em} with object size for three different periods: 
minimum, ascending and maximum phase of the current solar cycle. A correlation with solar activity for structures of 
size
$\sim$ 1000 ${\rm Mm}^{2}$ was reported,
with the highest fractal dimension being measured 
during the cycle maximum period.  Larger
structures (2000-7000 ${\rm Mm}^{2}$) were found to have a higher fractal 
dimension during the ascending phase of the cycle 
than at cycle maximum. Variations were of the 
order of few per cent. 
The same trends were found for estimates of \emph{D}, but with larger amplitude variations. 
If we restrict our analyses to the area range 2000-7000 ${\rm Mm}^{2}$, we instead find a little 
correlation of \emph{d1} with solar cycle, the highest
values being measured for years 2000 and 2002, and the smallest for 2005. 
The amplitude of the variations in our data is slightly higher than the one reported by \citet{Meunier2004}, 
the largest yearly variation measured 
over the six year period being of order 10\%. 
The trend reported for structures of moderate size (1000 ${\rm Mm}^{2}$) is not observed in our analysis.

\section{Discussion of fractal dimension estimation}

Assessment of the fractal dimension of features in digitalized images
requires a series of operations:
\begin{itemize}
\item Image segmentation to isolate regions of interest, 
\item Edge identification in the resulting bi-level images,
\item Perimeter and area measurement of structures so identified,
\item Fractal dimension evaluation using these measures.
\end{itemize}
Each of these steps introduces a certain degree of 
arbitrariness which influences the result. Moreover,
the results are sensitive
to intrinsic differences between image 
sets, unrelated to the geometric properties of the features they capture.
In this section we focus on the effects of edge identification technique, pixelization, and resolution,
by analyzing synthetic images of objects 
whose fractality is known: 
non-fractal objects, the von Koch curve,
and objects obtained by fractional Brownian motion.
Seeing effects are also investigated through
the analyses of MLSO PSPT data.

\subsection{Perimeter definition and pixelization effects} 
To study the influence of the perimeter finding algorithm, we examined the empirical dimension of
non-fractal objects as a function of their size. Three different perimeter identification techniques 
(described in \S2.4) were applied to three geometric shapes (squares, right triangles, and circles).
In the absence of error, all three methods should yield a value of one since the objects are non-fractal,
but because of image pixelization, fractal dimensions greater or lower than one were measured.

\begin{figure}[]
\includegraphics[width=7.0cm, height= 8 cm]{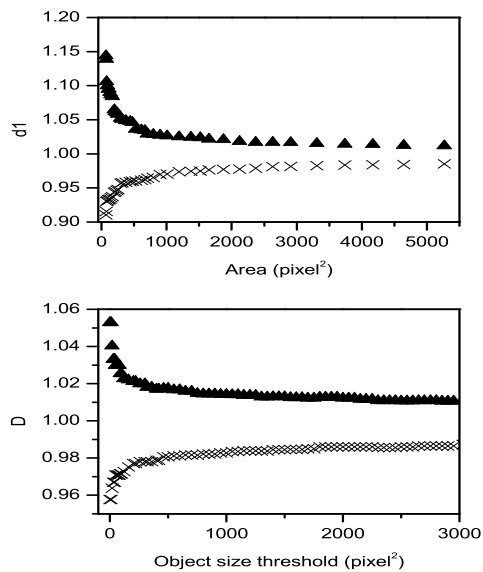}
\caption{Fractal dimension $d1$ (top) and $D$ (bottom) estimated for circles as function of object 
size and minimum area threshold respectively. Results obtained with external sides (crosses) and 8-contiguous point
(triangles) perimeter finding algorithm are shown. Error bars are smaller then symbol size.  Note that the horizontal scales 
differ for the two panels.
        \label{figcircled1dth}}
    \end{figure}

We found that errors in fractal 
dimension evaluation are functions of the object size for both \emph{D} and \emph{d1}.
Figure \ref{figcircled1dth} shows the results obtained for circular objects, with the top panel
showing \emph{d1} versus object area and the bottom 
panel plotting $D$, evaluated by fitting points
of area greater then a given threshold, as a function 
of the threshold value itself. In both cases, 
errors are greatest for objects of small size
but persist to surprisingly large scales. Analogous trends were observed for the other shapes analyzed.
In \emph{d1} estimations, errors of less than $5\%$ are achievable 
for object sizes greater than some hundreds-1000 pixel$^{2}$, but for circular objects,
which can not be grid aligned, 
the error never drops below $1\%$, independent of
the perimeter measure employed. This is true even for object sizes exceeding 5000 pixel$^{2}$.
For any given size object 
\begin{em}D\end{em} is significantly closer to its expected value of 1 
than is \begin{em}d1\end{em}. This is because the evaluation of \emph{D} in the perimeter-area fit
is performed over all points above a minimum size. This includes
the large objects not included at small scales in  
the evaluation of \emph{d1}. Therefore, the object size threshold above which the error in \emph{D} is below $5\%$ 
occurs at smaller scales than for \emph{d1}, but still is not usually less 
than some hundred square pixels.

The origin of these errors lies in the impossibility of representing curves or non-grid 
aligned lines on a rectangular grid. 
This causes the area and the perimeter to scale 
differently from what is expected for non
fractal objects. For instance, in the case of a right triangle whose two sides are grid aligned, the overestimation 
of the hypotenuse leads to the overestimation of both perimeter and area. It can be shown that, because the 
relative error in the perimeter
estimation is not size dependent, while the relative error in area estimation decreases with increasing
object size, the estimated
fractal dimension is always overestimated.

For our analysis of solar data, we employed only the row and 
column counting method (the first method described in \S2.4).
This method was chosen because it alone produced no error for 
grid aligned squares, a minimum criterion.

The analysis described above was also applied to fractal structures:
the von Koch snowflakes \citep{Peitgen1992} and fractional Browian 
motion (fBm) images \citep{fractgeomdigim}. For the first object, whose fractal dimension is $\sim 1.26$, we produced
snowflakes up to level 6 of different sizes (see Appendix) and studied their perimeter and area scaling. 
For fBms we created two sets of 150 images of expected 
fractal dimensions 1.8 and 1.6 respectively. Each fBm image was segmented with seven different thresholds \citep{fractgeomdigim}
and perimeter and area of the structures selected by the different thresholds were combined to study the fractal dimension.

\begin{table}  
\centering
\begin{tabular}{ccrr}                                        
\hline
$   $ & $D_{th}$  & \multicolumn{1}{c}{$D$}   &  \multicolumn{1}{c}{$\epsilon$}  \\
\hline
fBm & $1.8$     & $1.516$  &    $16 \%$ \\
fBm & $1.6$     & $1.413 $  &    $12 \%$ \\
vonKnoch & $1.26$    & $1.310$  &    $-3 \%$\\                        
\hline
\end{tabular}
\caption[]{Theoretical fractal dimension $D_{th}$, measured 
fractal dimension $D$ for the studied objects and 
the relative error $\epsilon=(D_{th}-D)/D_{th}$. Pixelization errors increase with
increasing structure complexity. 
  }\label{pixparset}
\end{table} 
\begin{figure}
\includegraphics[width=7 cm, height= 4.0 cm]{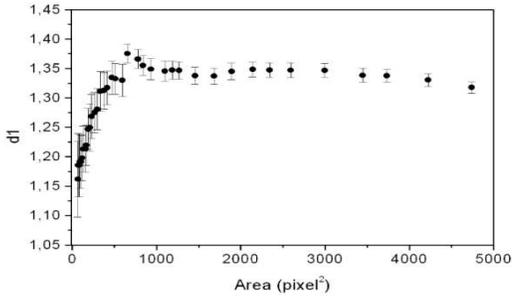}
\caption{ $d1$ evaluated for von Koch snowflakes of level 6. Like non fractal objects and real data, $d1$ increases
with object size and reaches a plateau at areas $\ge$ 1000 pixel$^{2}$. The plateau value, about 1.34, is an overestimate
of the snowflake fractal dimension (see text).
        \label{figlvl6}}
    \end{figure}
In fig.\ref{figlvl6}, the measured dimension $d1$ is plotted as a function of object size for 
von Koch snowflakes of level 6.
The fractal dimension increases with object size from a minimum value,
in this case 1.15, to an almost constant value approximating the theoretical one, over
the  size range 1000 to 5000 pixel$^2$.  The plot 
is surprisingly reminiscent of that found for 
real solar structures (\S3). Note that the plateau value, about 1.34,
 exceeds the one expected theoretically. This reflects the overestimation of the snowflake perimeter inherent
 in the perimeter measure algorithm employed, as discussed previously for simple non fractal triangle.   
A rise of fractal dimension with object size was also observed for fBm images. 
Measurements of \emph{D} are similarly affected by pixelization at small scales, with more complex
structures harder to resolve and thus showing greater 
measurement error in the deduced fractal dimension, as shown in table \ref{pixparset}. 
    
Both regular and fractal objects  
show similar pixelization induced errors in the fractal dimension estimation.
These effects are greater at smaller areas, 
where the lack of resolution
causes the objects to appear round, thus
both \begin{em}d1\end{em}
and  \begin{em}D\end{em} increase rapidly with object size for object areas 
less than $\sim500 - 1000$ pixel$^2$ and some hundred pixels square respectively.
For objects of larger area, \begin{em}D\end{em} and \begin{em}d1\end{em} 
increase more slowly, but show deviations from the
expected theoretical value reflecting how the structures map onto the pixel grid.

We thus suggest that the minimum object size thresholds (about some tens of
pixel$^2$) applied in previous works (e.g. \citealt{Vogelaar1994}) 
are 
insufficiently conservative, with residual pixelization effects significantly influencing the 
final results even for objects of $\sim1000$ pixel$^2$.

\subsection{Resolution and seeing effects}
The fractal dimension estimate of solar features depends on the 
resolution of 
the images analyzed. Resolution is determined not only by the detector pixel size (image scale), 
but also by the 
aperture of the telescope, any instrumental aberration, and, 
for ground based instrumentation,
the distortion introduced by atmospheric turbulence (seeing).
Thus  
the pixel scale and the resolution are not the same.
To evaluate the effects of resolution on the estimation of fractal dimension, 
we analyzed the scaling of 
\begin{em}d1\end{em} and \begin{em}D\end{em} with area after convolving von Koch snowflake
and fBm images with Gaussian functions of different widths.
We obtained, as one might have expected, a decrease in both \emph{d1} and \emph{D} 
accompanying the smoothing.
Table \ref{parset} of \emph{D} (fit over the entire 
perimeter area range) shows also that the smoothing effects become more important 
as the structure complexity increases.

\begin{table} %
\centering 
\begin{tabular}{ccrrc}                                        
\hline
$ $ & $D_{th}$  & \multicolumn{1}{c}{$D$}    & \multicolumn{1}{c}{$D_{sm}$} &  \multicolumn{1}{c}{$\epsilon$} \\
\hline
fBm & $1.80$     & $1.516 $  &  $1.366 $  &   $9.9 \%$\\
fBm & $1.60$     & $1.413 $  &  $1.303 $  &   $7.8 \%$\\
von Koch & $1.26$    & $1.310 $  &  $1.273 $  &   $2.8\%$\\                       
\hline
\end{tabular}
\caption[]{Fractal dimension measure for different fractals, 
before $D$ and after $D_{sm}$ smoothing by convolution with a Gaussian
of fwhm=2, and  the relative error $\epsilon=(D-D_{sm})/D$.  
Note that to distinguish resolution from pixelization induced effects, the error is
evaluated respect to $D$ and not to $D_{th}$.
 \label{parset}}
\end{table} 
\begin{figure}[]
\includegraphics[width=9cm]{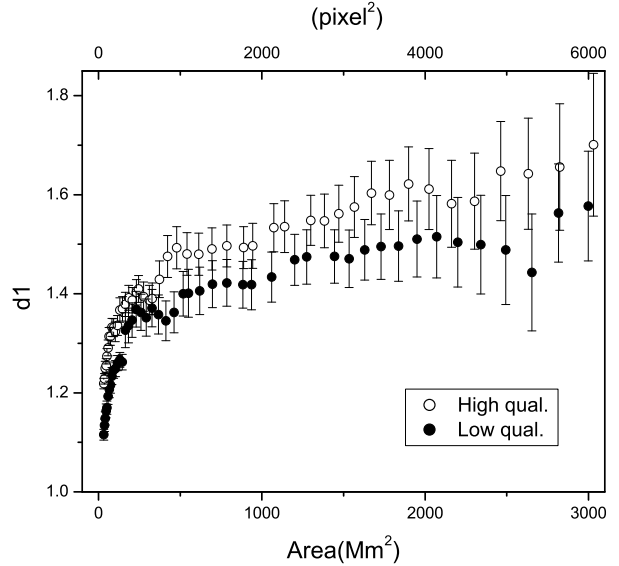}
\caption{Facular fractal dimension estimated on the two different full resolution MLSO quality sets 
described in the text. When the estimation is carried out on images less affected by seeing degradation, the measured 
fractal dimension is higher. 
        \label{figd1gbhawaii}}
    \end{figure}

A Gaussian function is a rough approximation to the seeing and instrumental aberration 
Point Spread Function in real images. Moreover, seeing is a time dependent phenomenon, so that images acquired
at different times are affected by different degradation. 
In order to investigate directly the effect of variable seeing on the computed fractal dimension of
structures in real data, we 
examined full resolution PSPT images from MLSO after selection based on 
the quality criteria described in \S2.2. Images were segmented with the first technique explained in \S2.3.
Figure \ref{figd1gbhawaii} shows that a real 
decrease in resolution resulting from degraded observing conditions 
leads to an underestimation of features' complexity at all scales.

\section{Interpretation of results measured}
We have shown that the determination of fractal dimension of features in digitalized images
by the two estimators   
\begin{em}d1\end{em} and \begin{em}D\end{em} is affected by pixelization 
and resolution. 
Understanding these effects is essential to the interpretation of the results obtained from
OAR and MLSO PSPT images (\S3) as well as those previously reported from studies carried out with 
similar techniques on other data. 
Pixelization errors occur at all scales, 
but are generally more important for the smallest objects. 
This causes the estimated fractal dimension to increase rapidly with object size and become
almost constant at areas larger than a critical threshold, that we estimated to be
$\sim 500 - 1000$ pixel$^2$ for \begin{em}d1\end{em} and some hundreds of pixel$^2$ for \begin{em}D\end{em}.
Seeing and instrumental induced image degradation smooths edges making structures appear rounder, resulting in a 
reduced fractal dimension. This effect is expected, on the basis of synthetic fractal data,
to be more important for more complex objects (\S5.2).

We thus suggest that the rise of \begin{em}d1\end{em} with object size observed in PSPT data, 
as well as for example in Meudon 
spectroheliograms \citep{nesme96} and MDI magnetograms \citep{Meunier1999,Meunier2004}, 
is  most likely
an effect of image pixelization, 
rather than a signature of an intrinsic multifractality of active regions. Conclusions drawn in \citet{Meunier2004} 
concerning a change in physical properties of magnetic structures at supergranular scales, should thus be reviewed
in light of the results shown in this paper. 
Pixelization is also likely the cause of the 'break of similarity' observed by \citet{Jansen2003},
the break occurring at the same pixel scale for both real and simulated images,
in spite of the different physical scale implied, and in the same area range suggested by our
synthetic fractal studies.

For larger objects, the fractal dimension estimate is most affected by
seeing. 
The values measured for large scale structures in the PSPT observations ($1.5$ to $1.7$) 
are therefore likely an underestimate of the real value. Nevertheless, in some cases 
pixelization can cause overestimation of the fractal dimension at largest areas, as shown for instance
for the von Koch snowflake. We cannot therefore in principle exclude 
some compensation
due to the combined effect of pixelization and reduction of resolution.
We finally note that MDI magnetograms, while not affected by seeing, 
are slightly defocused, so that the resolution is
twice that of the pixel scale \citep{Scherrer1995}. The same considerations made for results obtained with
ground based measurements
thus also apply to results obtained with MDI. 

A study of the fractal geometry of solar active regions and its variation with the magnetic activity
cycle is thus feasible if it focuses only on large features, employs a constant segmentation technique 
throughout, and utilizes data of consistently high quality.
The OAR PSPT images analyzed over a period of six years marginally meet these requirements.
They do not show, however, a clear correlation with the solar cycle.  
Moreover the variations measured in D appear to be dominated 
by variations in size distribution of the examined features
\citep{ermolli06}, which in turn weight 
the perimeter-area fit.

\section{Conclusions}
We have analyzed the fractal dimension  of bright features identified in solar chromospheric CaII K 
images obtained during the last six years, corresponding to the descending phase of solar cycle 23. 
The results obtained are in general agreement with those reported in literature,
in particular with those studies that have been carried out with similar fractal estimators 
(\begin{em}d1\end{em} and \begin{em}D\end{em}). 

We have also investigated the effects of pixelization and resolution on the fractal dimension estimates obtained, 
studying these effects on real and simulated data, the latter including both non-fractal 
objects and objects whose fractal properties are well known, von Koch snowflakes and fBm images.  
We have shown that fractal dimension estimates suffer from pixelization errors at all scales, but errors are generally 
more important for areas 
less than  $\sim500 - 1000$ pixel$^2$. Particularly, pixelization causes measured fractal dimensions
to increase with object size
in the case of both fractal and non-fractal objects. Our results thus indicate that the increase 
of fractal dimension with the feature size reported in literature
by some previous analyses 
is likely  an effect of pixelization and image degradation,
rather than a signature of an intrinsic multifractality of active regions.  
To reduce these effects, 
we thus suggest a restriction of the analyses to objects whose areas are larger then the quoted value.

Our analyses also showed that image degradation due to both seeing and instrumental effects smooths features 
edges making 
them appear rounder. 
The fractal dimension estimated is consequently lower than expected, even for
large objects.
Perhaps image degradation effects can be compensated for, as suggested in \citet{Jansen2003}.
Alternatively, a more careful estimate of these effects
can be carried out  
by the analyses of images of known complexity objects (eg. fBms) convolved with realistic Point Spread Functions.  
We did not apply either of these compensations to our data and leave them for future investigation.

Finally we note that, as demonstrated by \citet{Baveye1998} for the box counting  
method, pixelization effects can influence other fractal dimension  
estimators as well, and careful quantitative measure of the effects  
for each measure employed is essential to the
interpretation of the results. We have not addressed this problem in this work and leave also this issue
for future research.  


\begin{acknowledgements}
The authors acknowledge useful comments by J.K. Lawrence that refereed the article. The authors are also grateful
to D. Del Moro,
G. Consolini and H. Liu for fruitful discussions.    
\end{acknowledgements}

\newcommand{\ApJ}{ApJ}
\newcommand{\ApJL}{ApJ Lett.}
\newcommand{\AAp}{A\&A}
\newcommand{\JGR}{JGR}
\newcommand{\GRL}{GRL}
\newcommand{\MNRAS}{Mon. Not. Royal Astron. Soc.}
\newcommand{\PASP}{PASP}
\newcommand{\SPh}{Sol. Phys.}

\appendix 
\section{On the estimation of \emph{D} and \emph{d1} on digitalized images}

The estimators adopted in this paper to evaluate fractal dimension of selected structures are based on the 
perimeter-area relation. This consists of measuring the perimeter and area of a structure at different resolutions.
\begin{figure}[]
\includegraphics[width=8.5cm]{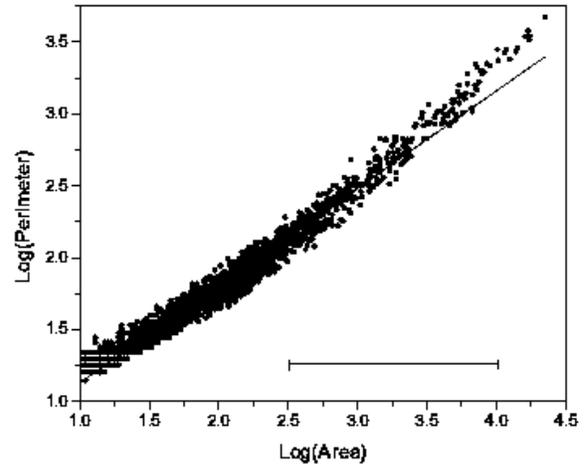}
\caption{ Perimeter (in units of pixel) and Area (in units of pixels square) in logarithmic scale of detected 
structures on
OAR PSPT data taken during summer 2002. Continuous line is the fit to the whole set of data ($D$=1.35).
Points at area larger than about 1000 pixels square are better approximated by a higher slope line. Horizontal line is the area window
width over which $d1$ is estimated.
        \label{rome02all}}
    \end{figure}
For regular structures,
perimeter scales as the square root of the area, while for fractal structures the exponent is greater than 0.5.
By definition, the exponent is the fractal dimension of the studied object.
Note that in \S2.5 we normalized the exponent so that the fractal dimension is one for regular structures and 
greater than one for fractal objects. It's worth noticing that, when adopting this method to investigate fractal nature of magnetic 
solar regions, 
one implicitly assumes that the different size selected structures are the 'same' object observed at different 
resolution.

\begin{figure*}
\centering
\includegraphics[width=9.5 cm ]{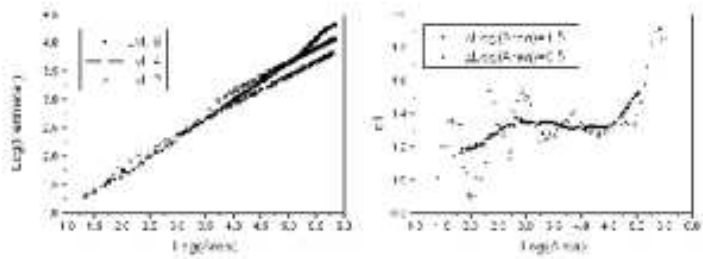}
\caption{Left: Perimeter versus area in logarithmic scale of snowflakes of levels 2,4, and 6. Because of pixelization,
these structures exhibit a fractal scaling 
only within certain area ranges, the 
bounds depending on the snowflake level. Right: $d1$ versus
area evaluated with different window sizes for snowflakes of level 6. Peaks obtained with the small window (open circles) 
are due to the steep variations 
visible in plot on the left. Peaks are not detected with a larger window (full circles). The area range over which 
$d1$ is almost constant is the range over which the simulated images are fractal. 
        \label{figareaperilvls}}
    \end{figure*}

When plotting the perimeter-area relation of real data, 
one expects to find at least three different regimes. At the smallest sizes, because of resolution, an object's detail is
is not fully detected, thus perimeter and area scale as for regular non-fractal structures. At the largest areas, a break 
in similarity can occur for 
physical reasons (for instance the object under study has a finite maximum size above which it is no longer
a fractal). At intermediate areas the object scales as a fractal.
As an example, in fig. \ref{rome02all} we plot the perimeter-area relation for structures selected in OAR-PSPT images.
The straight line is a linear fit to 
the points. It's evident that points don't lie on a single
line, but rather on a curve. One is thus tempted to measure the tangent to the curve as a 'local' measure of the 
fractal dimension.
This is what the measure in \emph{d1} employed in the text does. All our plots 
(cfr. figures \ref{figpspt},\ref{psptanni},\ref{figcircled1dth},
\ref{figlvl6},\ref{figd1gbhawaii}) showed that \emph{d1} increases with object size, becoming eventually
constant at some typical scale (generally in the area range 500-1000 square pixels). 
The measured \emph{d1}, as well as the area at which it becomes constant, are functions of the window size. 
Nevertheless, when using this estimator, it is important to keep in mind that 
any fractal estimate requires the 
autosimilarity to be valid over some 
orders of magnitude \citep{Baveye1998} (the area range $\Delta \log$ A =1.5 adopted for \emph{d1} in this and in other 
works is therefore 
in principle too small).

The results we obtained with von Koch snowflakes illustrate clearly these issues.
von Koch snowflake images of different sizes were produced 
following the iterative scheme 
of  \citet{Peitgen1992}. 
After each iteration, or level, the 
snowflake is more structured, with an increase in both perimeter and area. 
In the limit of infinite iterations,  
the perimeter tends toward infinity and the area approaches a finite value.
Here we investigate structures constructed with up to 6 levels.
The fractal dimension of the von Koch snowflake is $\log 4/\log 3 \approx 1.26$.\\ 
In left panel of fig. \ref{figareaperilvls}, the 
perimeter-area relationships for snowflakes of levels 2, 4, and 6 are plotted with
logarithmic scaling. 
For each level, the relationship traces a curve made 
up of segments whose slope is 1/2 connected by segments of slope greater than 1/2. 
At largest areas all the points lay on 
parallel lines of slope 1/2. At those scales the snowflakes of 
all the represented levels are fully resolved on the grid employed.
As the dimensions of the objects are reduced, fewer details at any fixed construction 
level are resolved, the measured perimeter decreases at a rate faster than $A^{1/2}$,
and the perimeter-area curve steepens. The slope flattens to a value of one-half again
each time the grid resolution is sufficient to capture the details of the next lower level.
Finally, at smallest areas most geometric details are lost and all the objects, independent
of their initial construction level, appear non-fractal.

The scaling of \begin{em}d1\end{em} better reflects
the change in
slope with objects size. As an example,
in right panel of fig.\ref{figareaperilvls} we show results obtained for level 6.
 Here full and open dots represent respectively \begin{em}d1\end{em} obtained with a window of $\Delta \log A=1.5$ and 
 a window of
 $\Delta \log A=0.5$. With the largest window only the slope change that occurs at largest areas is visible. 
 The others occur on scales smaller than the window so that they are not 'detected' and a plateau is observed. 
 At smallest areas \begin{em}d1\end{em} drops because of the resolution effects explained 
 before. When a smaller window is used, 6 peaks are visible, corresponding to the 6 slope-changes visible in the 
 perimeter area scatter plot.
In this case, there is an area range over 
which fractal dimension 
 oscillates around a constant value. At smallest areas larger amplitude oscillations are observed.
Both curves show clearly two of the three regimes mentioned
above. The object scales as a fractal in the range 
 3 $< \log$(Area) $<$ 4.5. 
At smaller areas, 
pixelization effects dominate the measurements because 
resolution is insufficient to allow
detection of all the structure's details. 
The third regime is not evident in the two curves, 
since not enough points
are available at largest areas to perform the fit. 
If more points were available we would observe a decrease 
of $d1$ toward the value of one, as slightly visible at largest areas for fits performed on the smaller window.
\\
  
\end{document}